# Paramagnetic Phase of a Heavy-Fermion Compound, CeFePO, Probed by $^{57}$Fe Mössbauer Spectroscopy

Tetsuro Nakamura[1], Takashi Yamamoto[2], Masanori Matoba[1], Yasuaki Einaga[2], Yoichi Kamihara[1,3]

[1]*Department of Applied Physics and Physico-Informatics, Faculty of Science and Technology, Keio University, Yokohama 223-8522*

[2]*Department of Chemistry, Faculty of Science and Technology, Keio University, 3-14-1 Hiyoshi, Yokohama 223-8522*

[3]*TRIP, Japan Science and Technology Agency, Sanban-cho bldg, 5, Sanban-cho, Chiyoda, Tokyo 102-0075*

$^{57}$Fe Mössbauer spectroscopy was applied to an iron-based layered compound CeFePO. At temperatures from 9.4 to 293 K, no magnetic splitting was observed in the Mössbauer spectra of CeFePO indicating a paramagnetic phase of the Fe magnetic sublattice. All the spectra were fitted with a small quadrupole splitting, and the Debye temperature of CeFePO was found to be ~448 K. The isomer shift at room temperature, 0.32 mm/s, was almost equal to those of *Ln*FeAsO (*Ln* = La, Ce, Sm). Comparing s-electron density using the isomer shifts and unit cell volumes, it was found that the Fe of CeFePO has a similar valence state to other layered iron-based quaternary oxypnictides except LaFePO.

## 1. Introduction

Since the unexpected discovery of a set of superconductors with an iron-based layer, LaFePO[1] and LaFeAsO[2], quaternary rare-earth transition oxypnictides with the chemical formula $Ln$Fe$Pn$O ($Ln$ = rare earth metal, $Pn$ = P, As, Sb, Bi) have attracted much attention. Because their family may have a different mechanism of superconductivity from superconducting cuprates, further systematic investigation of $Ln$Fe$Pn$O is crucial to clarify the mechanism of superconductivity and to further discover superconductors with high-$T_c$. The crystal structures of $Ln$Fe$Pn$O are identical at room temperature, belonging to the tetragonal P4/nmm space group. The unit cell contains two molecules and the chemical formula is represented by ($Ln_2O_2$)(Fe$_2Pn_2$), forming a layered crystal structure composed of an alternating stack of the ($Ln_2O_2$) and (Fe$_2Pn_2$) layers. It is suggested that superconducting transition temperatures reach their maximum for regular Fe$Pn_4$ tetrahedrons.[3]

CeFePO is a heavy-fermion material with the Ce 4$f$-electron's sublattice inducing Kondo effects. It has been reported that the Kondo temperature and the Sommerfeld coefficient of CeFePO are $T_K \sim 10$ K and $\gamma = 700$ mJ/mol K$^2$, respectively.[4] The Sommerfeld coefficient is about 56 times as large as that of LaFePO ($\gamma = 12.5$ mJ/mol K$^2$) indicating a strong Ce 4$f$-electron correlation. Magnetic ordering is almost quenched and/or suppressed in CeFePO, but at the metamagnetic field $H_M \sim 4$ tesla, the Kondo state breaks down and CeFePO exhibits a metamagnetic behavior.[5] Although its neighbor LaFePO undergoes a superconducting transition at ~5 K and CeFeAsO shows high-$T_C$ up to 35 K by F-doping,[6,7] no superconducting transition has been observed in CeFePO.

Although many studies of electrical transport properties, magnetic properties, and specific heat have been performed on CeFePO polycrystalline samples, only a $^{31}$P nuclear magnetic resonance (NMR) study has been reported as a microscopic measurement for FeP layers in CeFePO.[4-6] In this study, we will demonstrate the element-specific magnetism and valence state information of the Fe sublattice in CeFePO using $^{57}$Fe Mössbauer spectroscopy. Since $^{57}$Fe Mössbauer spectroscopy works on $^{57}$Fe nuclear, it enables us to make direct observation of the Fe magnetic phase and its strength of covalency. Both of them are necessary to discuss the electronic behavior near the Fermi surface, because in $Ln$Fe$Pn$O, the bands around the Fermi level are mainly formed by Fe 3$d$-states.

## 2. Experimental Procedure

A CeFePO polycrystal was synthesized by the method described in ref. 6. In addition, to serve as the FeAs-layered reference, CeFeAsO was prepared in the subsequent step.

Polycrystalline samples were prepared by a two-step solid-state reaction in a sealed silica tube, using dehydrated $CeO_2$ and a mixture of compounds composed of CeAs and $Fe_2As$ (CeAs-$Fe_2As$ powder) as starting materials. The dehydrated $CeO_2$ was prepared by heating commercial $CeO_2$ powder (Kojundo Chemical Laboratory; 99.99 wt.%) at 800 °C for 5 h in air. To obtain CeAs-$Fe_2As$ powder, Ce (Nilaco; 99.9 wt.%), Fe (Kojundo Chemical Laboratory; >99.9 wt.%), and As (Kojundo Chemical Laboratory; 99.9999 wt.%) were mixed in a stoichiometric ratio of 1:2:2 and heated at 880 °C for 10 h in an evacuated silica tube. Then, a 1:1 mixture of the two powders (dehydrated $CeO_2$ and CeAs-$Fe_2As$ powders) was pressed and heated in a sealed silica tube at 1310 °C for 15 h to obtain a sintered pellet. To prevent the silica tube from collapsing during the reaction, the tube was filled with high-purity Ar gas with a pressure of 0.2 atm at room temperature (RT). All procedures were carried out in an Ar-filled glove box (MIWA Mfg; $O_2$, $H_2O$ < 1 ppm).

The experiment on the two compounds, CeFePO and CeFeAsO, was performed with conventional $^{57}Fe$ Mössbauer equipment using 14.4 keV gamma-rays from a $^{57}Co$ source in an Rh matrix. Each spectrum was calibrated using an α-Fe foil. The velocity scale, that is, the isomer shift, was referenced to α-Fe. The measurements at 9.4, 50, and 293 K were carried out using a thermal shield cryostat without any vibrations, while the measurements at 100 and 200 K were performed under vacuum purging to prevent the cryostat from sweating. The spectra were fitted to Lorentzian line shapes using the MossWin[8] program.

### 3. Results

Figure 1 shows the Mössbauer spectra of CeFePO at 9.4, 50 100, 200 and 293 K. All the spectra were fitted with no magnetic splitting indicating the paramagnetic phase of the Fe sublattice. Corresponding fitting parameters are listed in Table I. The spectra were subjected to a weak quadrupole splitting indicating a distorted $FeP_4$ tetrahedron. The Mössbauer spectrum of CeFeAsO at 294 K was fitted to a singlet pattern, indicating that the quadrupole splitting was zero within the error bar (Fig. 2). The value of quadrupole splitting ($\Delta E_Q$) of CeFePO roughly increases with decreasing temperature. The enhanced line widths ($\Gamma$) at 100 and 200 K are not an intrinsic property of CeFePO but due to small vibration of the equipment due to the vibrations originating from vacuum purging, but little do they affect the isomer shift or quadrupole splitting as discovered bellow.

Figure 3 shows the temperature dependences of the isomer shift $\delta(T)$ of the CeFePO Mössbauer spectrum. In general, $\delta(T)$ is given by

$$\delta(T) = \delta_0 + \delta_{SOD}(T), \qquad (1)$$

where $\delta_0$ is the intrinsic isomer shift and $\delta_{SOD}(T)$ is the second-order Doppler shift that depends on the lattice vibrations of the Fe atoms. In terms of the Debye approximation of the lattice vibrations, $\delta_{SOD}(T)$ is expressed by the Debye temperature ($\Theta_D$) as

$$\delta_{SOD}(T) = -E_\gamma \times \left\{ \frac{3}{2} \frac{k_B T}{Mc^2} \left[ \frac{3}{8} \frac{\Theta_D}{T} + 3 \left( \frac{T}{\Theta_D} \right)^3 \int_0^{\Theta_D/T} \frac{x^3}{e^x - 1} dx \right] \right\}, \qquad (2)$$

where $E_\gamma$ is the energy of gamma-rays, $M$ is the mass of the Mössbauer nucleus and $c$ is the speed of light.[9] The quantities $\delta_0$ and $\Theta_D$ were found to be 0.55(0) mm/s and 448(31) K, respectively, by fitting the experimental data to eqs. (1) and (2),.

## 4. Discussion

Figure 4 shows the isomer shift ($\delta$) of various layered compounds at room temperature as a function of the inverse of unit-cell volume.[10-13] Details are listed in Table II. Since $\delta$ depends on $s$-electron densities at the nucleus of the absorber and emitter, $|\psi(0)_a|^2$ and $|\psi(0)_e|^2$, as

$$\delta = \frac{2}{5} \pi Z e^2 \left( R_e^2 - R_g^2 \right) \left( |\psi(0)_a|^2 - |\psi(0)_b|^2 \right), \qquad (5)$$

where $R_e$ and $R_g$ are the radii of the excited state and ground state,[14] respectively, and $\delta$ is an indicator of covalency. Therefore, in Fig. 4, the situation in which the plots are roughly located linearly indicates the similarity of the covalency between the Fe-$Pn$ layer in CeFePO and of that in the other layered compounds. On the other hand, the $\delta$ of LaFePO is relatively small, indicating that LaFePO has a rather strong Fe-$Pn$ covalency and/or the effect of the Fe-Fe bond is being enhanced in LaFePO. In either case, the strong covalency and hybridization effects of LaFePO increase its bandwidth for Fe 3$d$-states, resulting in a better metallic behavior with weaker electronic correlations.[15-17] This fact is further supported by the band calculations by Vidosola $et\ al.$[18] In addition, the specifically strong covalency might be related to the uncertainty of the $T_C$ of LaFePO being under controversy.[15-17]

The quadrupole splitting ($\Delta E_Q$) of CeFePO at room temperature, 0.228 mm/s, is much larger than those of CeFeAsO (0.00 mm/s), LaFeAsO (0.03 mm/s), and other FeAs-layered systems.[10,19] Quadrupole splitting indicates the existence of an electric field gradient due to a distorted FeP$_4$ tetrahedron, and the appearance of Jahn-Teller splitting. The Jahn-Teller splitting of $d$-orbital-derived bands increases the bandwidth, which results in relatively itinerant *d*-electrons [Fig. 5].[17] Providing that superconductivity is enhanced by increasing correlation strength ($U/W$ in Fig. 5), this is one of the reasons why CeFePO is unsuitable as a mother material of high-*T*c superconductors compared with *Ln*FeAsO.

## 5. Conclusions

An $^{57}$Fe Mössbauer spectroscopy measurement of CeFePO was performed. The Fe in CeFePO showed no magnetic order down to 9.4 K. The Debye temperature of CeFePO was found to be ~448 K. The Fe of CeFePO has almost the same valence state as the other layered compounds except LaFePO. CeFePO exhibits a larger quadrupole splitting value ($\Delta E_Q$) than CeFeAsO. A finite $\Delta E_Q$ indicates the appearance of Jahn-Teller splitting. The Jahn-Teller splitting of the *d*-orbital-derived bands increases the bandwidth, which results in relatively itinerant *d*-electrons CeFePO.

**Acknowledgments**

Y. K. is indebted to Drs. H. Hosono (Tokyo Institute of Technology), K. Ishida, and M. Seto (Kyoto University) for their helpful advice. This work was partially supported by the research grants from Keio University, the Keio Leading-edge Laboratory of Science and Technology (KLL), and Funding Program for World-Leading Innovative R&D on Science and Technology (FIRST) from Japan Society for Promotion of Science (JSPS).

Fig. 1. (Color Online) Observed $^{57}$Fe Mössbauer spectra of CeFePO at 293, 200, 100, 50, and 9.4 K. Each solid line is a theoretical spectrum that includes a finite quadrupole splitting. Two broad spectra at 100 and 200 K are measured under slight vibrations that are due to vacuum purging to prevent the cryostat from sweating.

Fig. 2. (Color Online) Observed $^{57}$Fe Mössbauer spectrum of CeFeAsO at 294 K. The solid line is a theoretical spectrum composed of a singlet. The quadrupole splitting wass zero within the error bar.

Fig. 3. (Color Online) Temperature dependence of isomer shift ($\delta$) of the $^{57}$Fe nucleus in CeFePO. The open circles are the experimental points; the broken lines are the least-squares fit to eqs. (1) and (2) (see text).

Fig. 4. (Color Online) Isomer shift of $Ln$Fe$Pn$O[10-13] as a function of the inverse of the unit-cell volume at room temperature. The isomer shift of LaFePO is relatively small, indicating that LaFePO has a rather strong Fe-$Pn$ covalency and/or that the effect of Fe-Fe bond is enhanced in LaFePO.

Fig. 5. (Color Online) (a) Splitting of the energy levels of the Fe 3$d$-orbitals by crystal field. The Jahn-Teller splitting ($\varepsilon$) appears in a distorted tetrahedron. (b) Simplified orbital correlation diagram for Fe 3$d$ and $Pn$ 3$p$ in $Ln$Fe$Pn$O. Since the Jahn-Teller splitting ($\varepsilon$) increases the bandwidth ($W$), $\varepsilon$ is a factor for the itinerant nature of $d$-electrons in $Ln$FePO.

Table I. Fitting parameters of isomer shift ($\delta$), quadrupole splitting ($\Delta E_Q$) and line width ($\Gamma$) for $^{57}$Fe Mössbauer spectra of CeFePO.

| $T$(K) | $\delta$(mm/s) | $\Delta E_Q$(mm/s) | $\Gamma$(mm/s) |
|---|---|---|---|
| 293.0(1) | 0.320(1) | 0.136(5) | 0.354(5) |
| 200.0(1) | 0.368(2) | 0.203(11) | 0.606(9) |
| 100.0(1) | 0.423(1) | 0.228(7) | 0.564(7) |
| 50.0(1) | 0.429(1) | 0.217(4) | 0.446(5) |
| 9.4(1) | 0.434(1) | 0.228(5) | 0.465(6) |

Table II. Unit-cell volume ($V$) and isomer shifts ($\delta$) at room temperature of $Ln$Fe$Pn$O[10-13] ($Ln$ = La, Ce, Sm; $Pn$ = P, As).

| Compound | $V$ (nm$^3$) | $\delta$ (mm/s) | $T$ (K) | Reference |
|---|---|---|---|---|
| LaFeAsO | 0.1422 | 0.45 | 298 | Kitao *et al.* [10] |
| CeFeAsO | 0.1390 | 0.43 | 294 | Present work |
| SmFeAsO | 0.1374 | 0.40 | 298 | Kamihara *et al.* [11] |
| LaFePO | 0.1336 | 0.24 | 298 | Tegel *et al.* [12] |
| CeFePO | 0.1280 | 0.32 | 293 | Present work |
| SmFePO | 0.1236 | 0.28 | 298 | Kamihara *et al.* [13] |

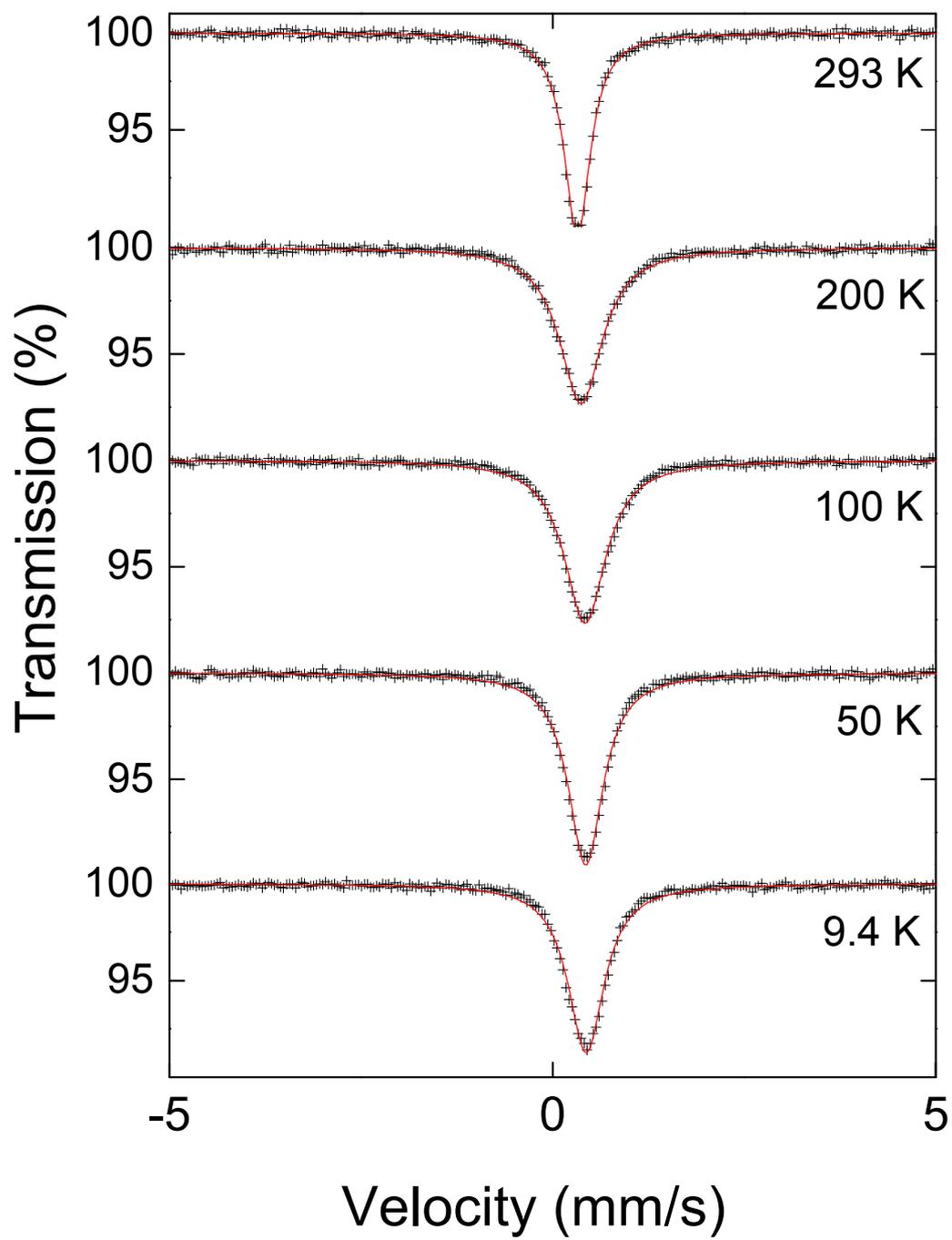

Figure 1.

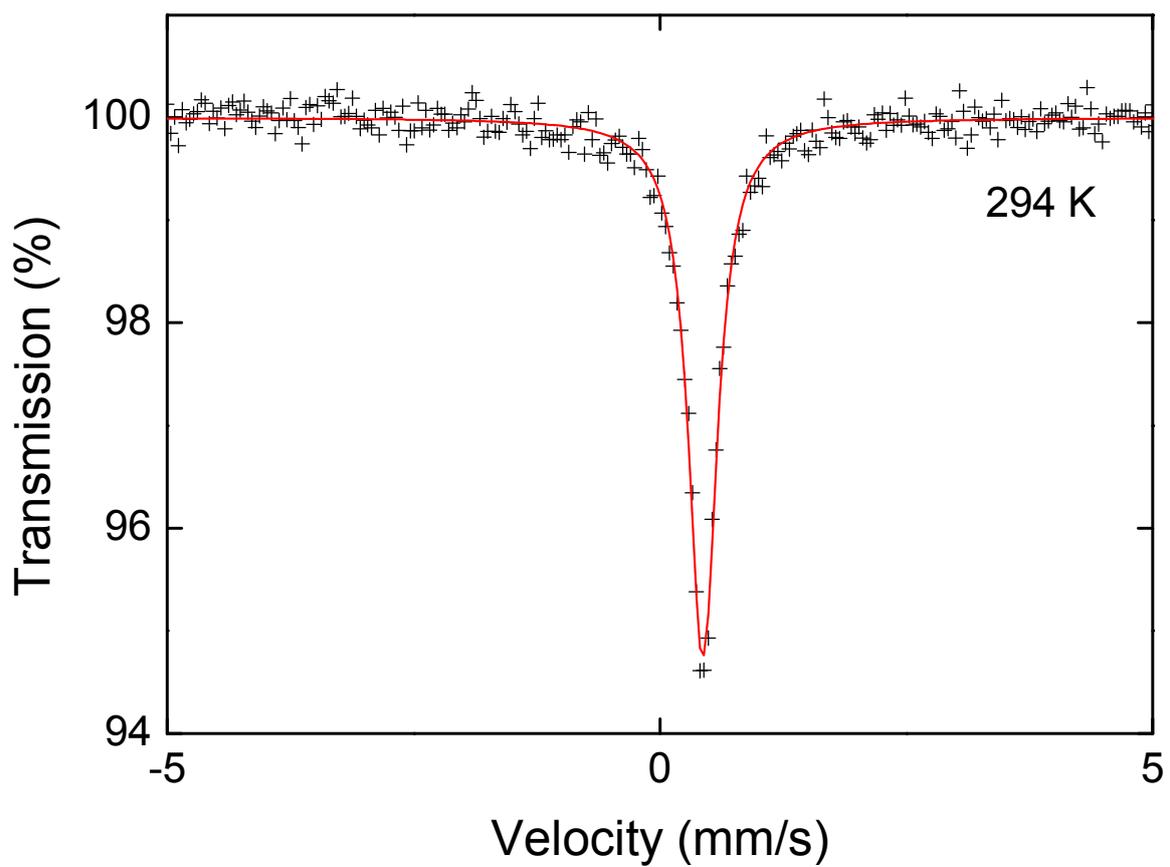

Figure 2.

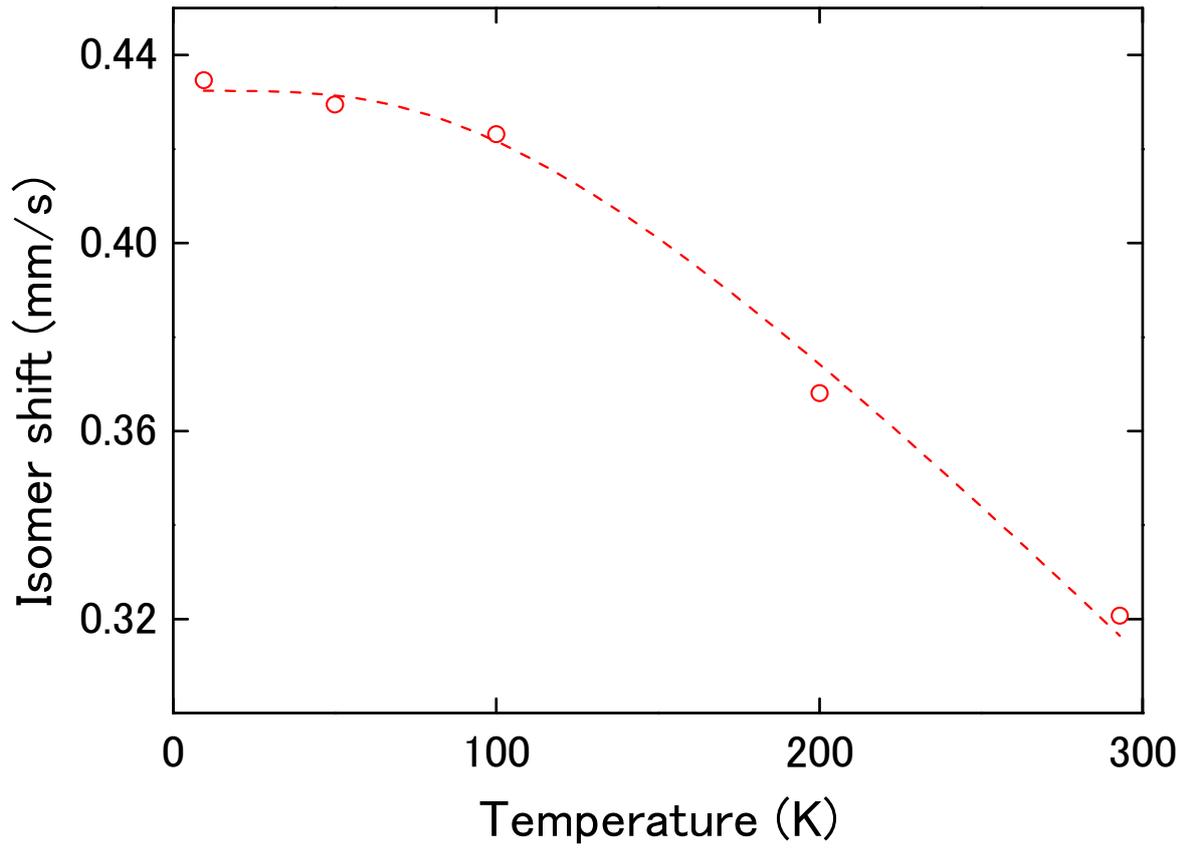

Figure 3.

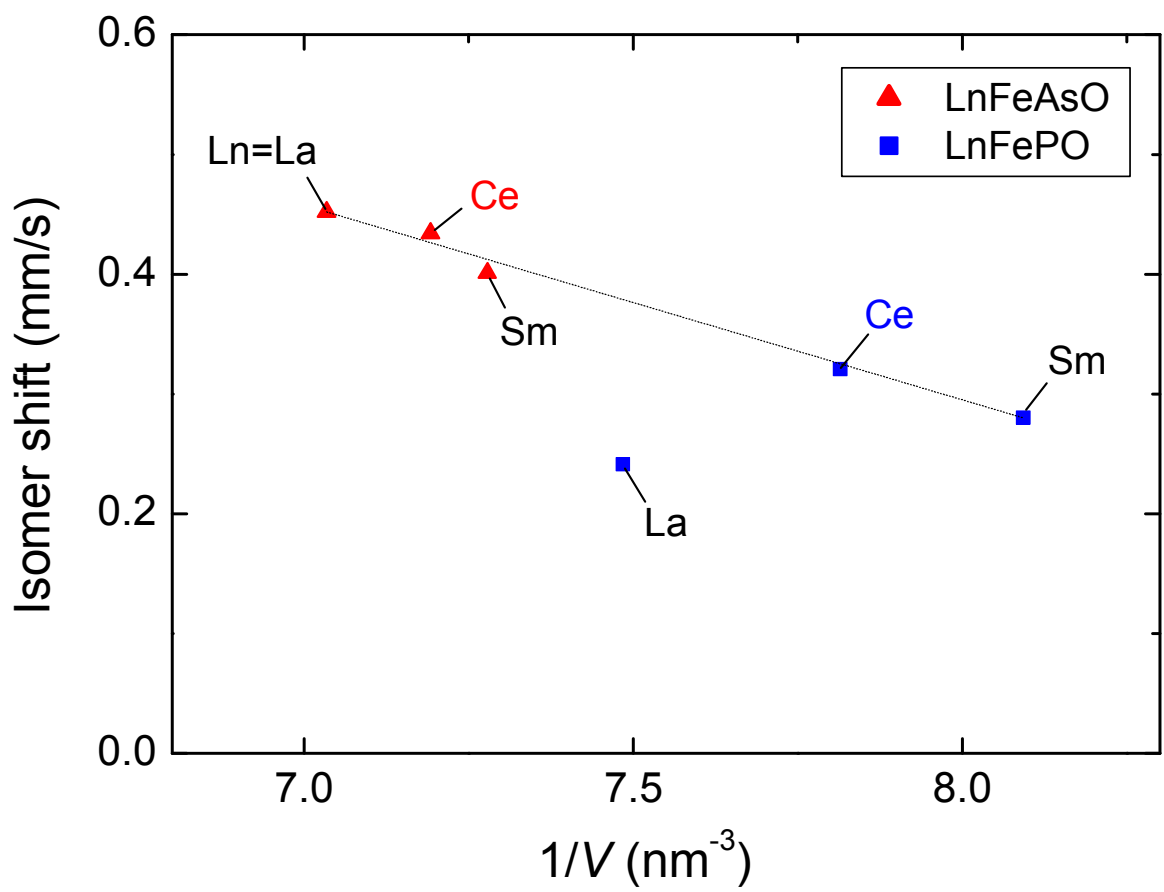

Figure 4.

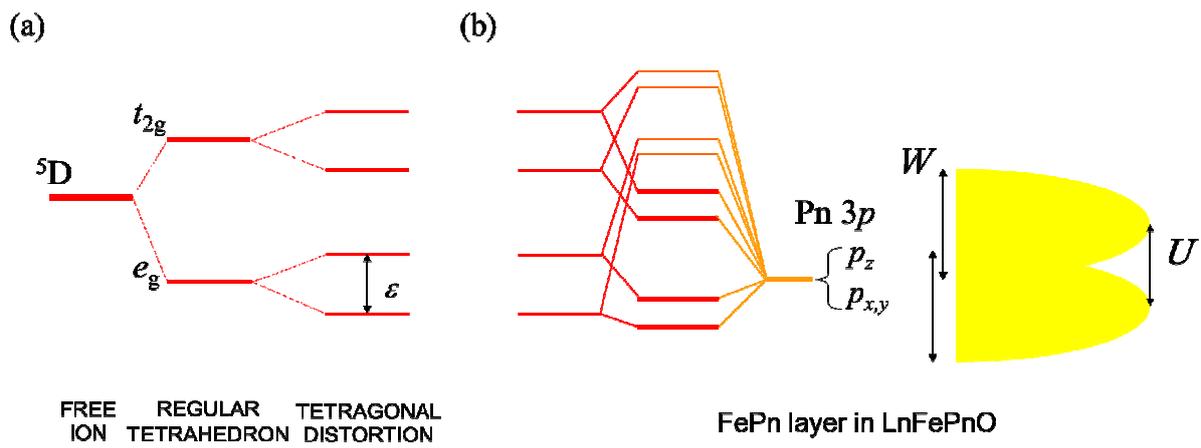

Figure 5.